# Electric-field Controlled Magnetization Switching in Co/Pt thin-Film Ferromagnets

A. Siddique[1], S. Gu[1], R. Witte[2], M. Ghahremani[1], C. A. Nwokoye[1], A. Aslani[1], R. Kruk[2], V. Provenzano[1], L. H. Bennett[1], and E. Della Torre[1]

[1]Department of Electrical and Computer Engineering, George Washington University, Washington, DC, USA

[2]Institute for Nanotechnology, Karlsruhe Institute of Technology, Karlsruhe, Germany

**Abstract:** A study of dynamic and reversible voltage controlled magnetization switching in ferromagnetic Co/Pt thin film with perpendicular magnetic anisotropy at room temperature is presented. The change in the magnetic properties of the system is observed in a relatively thick film of 15 nm. A surface charge is induced by the formation of electrochemical double layer between the metallic thin film and non-aqueous lithium $LiClO_4$ electrolyte to manipulate the magnetism. The change in the magnetic properties occurred by the application of an external electric field. As the negative voltage was increased, the coercivity and the switching magnetic field decreased thus activating magnetization switching. The results are envisaged to lead to faster and ultra-low power magnetization switching as compared to spin-transfer torque (STT) switching in spintronic devices.

The shift from the longitudinal to perpendicular mode in the magnetic media in the last few years has seen an enormous growth in Areal Density (AD). Currently, the magnetic recording industry is striving to push the AD of stored information beyond the superparamagnetic limit of 1 $Tbit/in^2$. The progress is challenged by the superparamagnetic trilemma [1], which becomes significant at smaller bit size. Manipulation of the magnetization direction of the magnetic bit via electric current induced magnetic field has been used [2-4]. However, as the bit size is miniaturized, the required electric current increases dramatically. Even though the required current reduces as the bit size is reduced, the electric current is still too large for practical applications.

Voltage driven modulation is proposed as a desirable alternative method for electrical control of magnetic properties, which can circumvent the limitations in the other methods, as it does not require electric current flow. Recently, the use of electrostatic fields applied at a solid-solid [4-14] or liquid-solid [15] interface to achieve dynamic and reversible control of material properties (tuning) has gained momentum. Such a method to control magnetic properties, via application of an electric field-effect gating principle, well known in transistors, is of fundamental interest in any application concerned with the manipulation, storage, and transfer of information by means of electron spin. It has been shown that for many nanostructures an applied electrostatic field, or surface charge, can bring about a noticeable magneto-electric response for various ferro- and ferrimagnetic materials [16-18]. As a spectacular experiment benchmark one can consider reversible control over magnetic phase state in metals, which has been demonstrated in ultra-thin films (a few atomic monolayers) of Pt/Co, Fe upon dielectric and electrochemical charging [5,20].

The $CoPt_3$ film has strong perpendicular magnetic anisotropy (PMA) and is a type of materials of interest for the magnetic recording industry and is used here. The response of the magnetization to electrochemical surface charging by the non aqueous electrolyte was studied using a MOKE magnetometer at room temperature. The most significant difference between the $CoPt_3$ film reported here and those in [5, 20] is the film thickness. We used a much thicker film, in the range of 15 nm. Yet, despite the thickness, the magnetic anisotropy of the whole film is affected by an applied surface charge. Normally, in metals a large carrier concentration limits the screening length of the external field to a few monolayers. This implies a new mechanism is necessary for the thicker sample.

*Sample Synthesis:* Co/Pt sample shown in Fig. 1 was provided by R.F.C. Farrow (IBM Research division, Almaden Research center). The Rutherford backscattering spectrometry (RBS) performed on the sample revealed the composition of the sample as shown in Fig 2. The darker area in the film is Si substrate and the $CoPt_3$ thin film with <101> texture and (111) reflectivity having thickness of 15nm was grown with perpendicular magnetic anisotropy in shiny circles. The film was capped with 5nm of Pt to stabilize against the surface oxidization.

*Experimental Setup:* We designed and built an *in-situ* experimental system for the measurement of surface charge induced variations in magnetic properties, which consists of a MOKE apparatus, a three-electrode electrochemical cell and a Gamry G300/750 potentiostat as shown in Figure 3. The electrochemical consists of three electrodes, working electrode, i.e. the sample, counter electrode (Pt), a reference electrode (Ag/AgCl), and a Teflon cap. A 1M Lithium Perchlorate (non-aqueous electrolyte, 99% pure, Sigma Aldrich) dissolved in ethyl acetate (non-aqueous solution, Sigma Aldrich) was chosen to provide a stable wider electrochemical potential window with a negligible electrolysis of electrolyte in this range of applied voltage. For the ferromagnetic Co/Pt ultrathin film sample that exhibits perpendicular magnetization, the MOKE apparatus is set to operate in the polar mode. The electrochemical cell is positioned between the two electromagnets and aligned so that the incident and reflection laser beam can directly pass through a small hole in the pole cap of the electromagnet so that both the incident and the reflection laser beam are almost perpendicular to the surface of the working electrode. The potentiostat is used to accurately control the potential of the counter electrode (CE) with respect to that of the working electrode (WE). The potential is measured with respect to the Ag/AgCl reference electrode (RE). The working electrode is charged electrochemically when a non-zero bias potential between the counter electrode and the working electrode is applied under potentiostatic control. When a magnetic field is applied, the magnetic response of the sample under various user-specified values of bias potential are measured using the MOKE magnetometer.

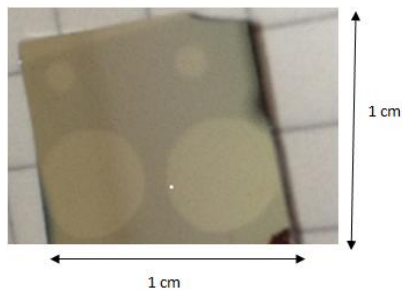

Fig. 1 Image of CoPt$_3$ Sample

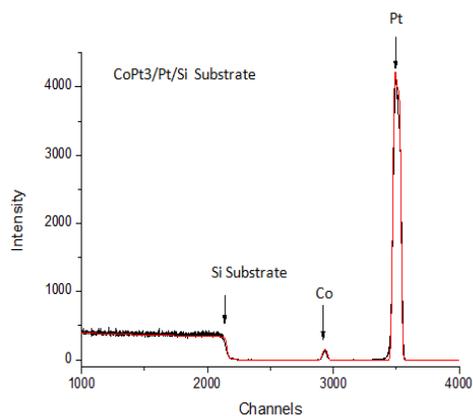

Fig. 2 Rutherford Backscattering Spectrum (RBS) of CoPt$_3$

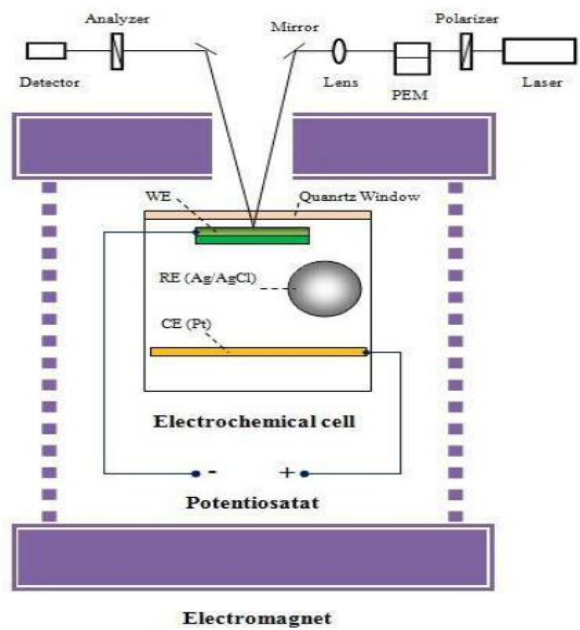

Figure 3: Schematic illustration (top view) of the in-situ experimental setup that consists of a MOKE apparatus, a potentiostat and a three-electrode electrochemical cell

Even though the electric field can alter the spin polarization at the surface of the solid nanostructure using dielectric gating, in the case of metals, the change in charge carrier concentration can hardly be detected, due to the extremely small screening length which is a few monolayers. In comparison, the surface charge that can be achieved using electrochemical gating via the formation of electrolyte double layer is up to two orders of magnitude higher than the maximum achievable polarization using dielectric gating. Hence, significant electric field induced changes in macroscopic properties of metallic nanostructure can be expected. Note, it should be emphasized that electrochemical charging does not require an insulating dielectric layer and therefore a large amount of charge can be stored at low voltages. In addition, wider classes of nanostructures can have access to the surface charging using electrochemical gating, where the electronic conductivity is measured *in situ* as a function of the charge carrier density as it is varied by the electrochemical potential. When the solid/dielectric interface is used for surface charge, the interface is required to be completely smooth and defect free, imposing a challenge in fabrication.

Cyclical voltammetry was carried out to ensure the stable electrochemical potential window. The electrochemical window was found to be 4.27 V.

$E = \varphi^+ - \varphi^- = 4.27V$

where

$\varphi^+ = \varphi\ ClO_4^-/ClO_3^- = 1.23\ V$

$\varphi^- = \varphi\ Li^+/Li = -3.04\ V$

As it is shown in the Figure 4, a voltage sweep between +1.3 V and -0.9 V was scanned at 30mV/s across the electrodes, and there was significant electrochemical charging taking place at the working electrode.

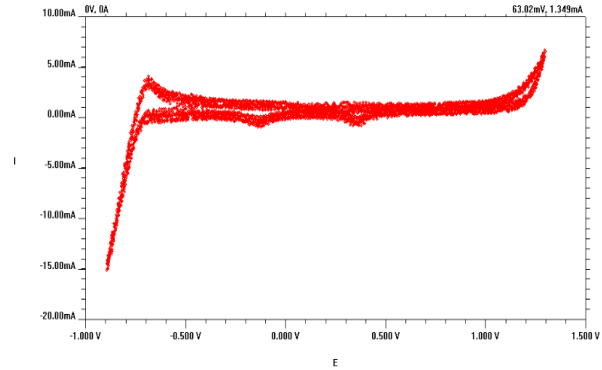

Figure 4: Cyclic Voltammetry Plot for voltage sweep between +1.3V and -0.9V

*Experimental Procedure:* An experimental procedure is used in the present study to systematically evaluate the effect of induced surface charge on critical magnetic parameters. We measured the hysteresis loop (Polar Kerr rotation vs. applied magnetic field) without potentiostatic control, i.e., the gate voltage, $V_G = 0$. Then we set the gate voltage to a non-zero value of interest within the potential window (for electrochemical double layer formation) and measure hysteresis loop. Thereafter, we calculated and analyzed the percentage variation of critical magnetic parameters, e.g. coercivity with the application of gate voltage. Afterwards, the same procedure was repeated with different bias voltages.

*Results:* Figure 5 shows the observed response of polar Kerr rotation to the application of negatively biased voltage of -0.5 V. The sample was first saturated with the large negative magnetic field under zero gate voltage. Then, the magnetic field was increased to point "A", denoted by a black arrow, and a gate voltage of -0.5 V was induced to zero at point A. The nucleation of the reversed domain has taken place at "A". The subsequent magnetization switching follows the blue solid curve, instead of the black dashed curve, as the magnetic field increased to positive saturation. Thereafter, the magnetic field was decreased to zero at point "B", denoted by a blue arrow, and the gate voltage at B was removed. The subsequent magnetization switching follows the black solid curve, instead of the blue dashed curve, to negative saturation.

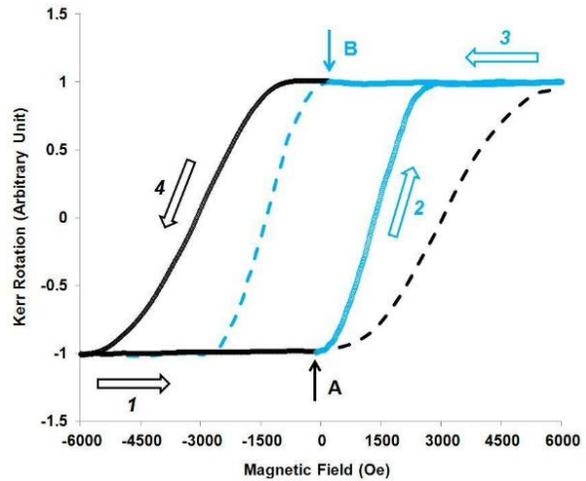

Figure 5: Dynamic control of magnetization switching via gate voltage application. A gate voltage of -0.5 V is induced at A, denoted by black arrow, and then removed at B, denoted by blue arrow.

When different negative gate voltages are applied at room temperature as shown in Figure 6, the Co/Pt system exhibits different ferromagnetic behavior in hysteresis curves. When the gate voltage changed from 0 V to -0.8 V, that is, the magnitude of negative gate voltage increased, significant changes in critical magnetic parameters were observed; the saturation magnetization, coercivity, and switching field distribution decreased, however, the perpendicular anisotropy clearly increased.

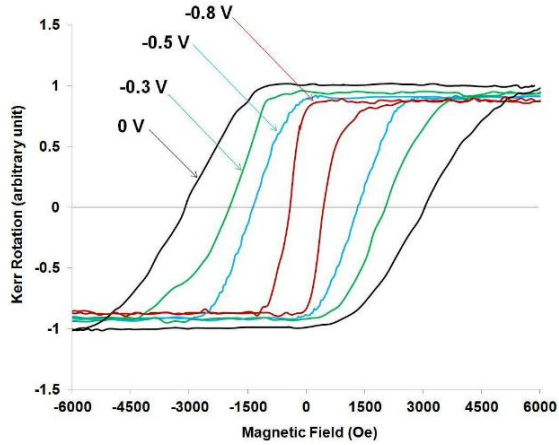

Figure 6: Polar Kerr hysteresis loops at 300 K under gate Voltages 0 V,-0.3 V, -0.5 V and -0.8 V. Hysteresis curves are normalized with respect the saturation under the gate voltage of 0 V.

The change in coercivity, in particular, indicates that magnetization switching process can be modulated electrically. The magnetization reversal process starts with the nucleation of small reversed domains, followed by the subsequent domain wall propagation. The change in domain wall propagation results in modulated activation energy barrier, and therefore leads to the variation in coercivity. The change in coercivity, $\Delta H_c$, as a function of gate voltage is shown in Figure 7.

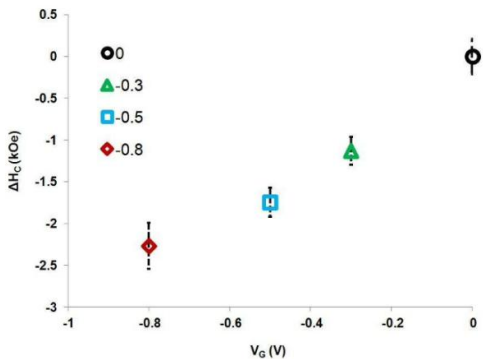

Figure 7: Change in coercivity as a function of gate voltage. The coercivities under various gate voltages are obtained from the hysteresis curves.

The application of a sufficiently large negative gate voltage, which leads to significant reduction of coercivity and large induced perpendicular anisotropy, may also be used to trigger the magnetization switching to a stable magnetization state. As shown in Figure 8, a positive saturation field is first applied to the Co/Pt system magnetic field under zero gate voltage, the field is then decreased and fixed at "C", a value indicated by a vertical dashed line, at which the magnetization state is stable under zero gate voltage. It should be noted that the magnitude of "C" is larger than that of the coercivity under a gate voltage application of -0.8 V. That means, when a gate voltage of -0.8 V is applied, the magnetization state at "C" can no longer be sustained and magnetization switching will be immediately activated. The magnetization switching can therefore be dynamically controlled by inducing a gate voltage at various stages of the magnetization reversal.

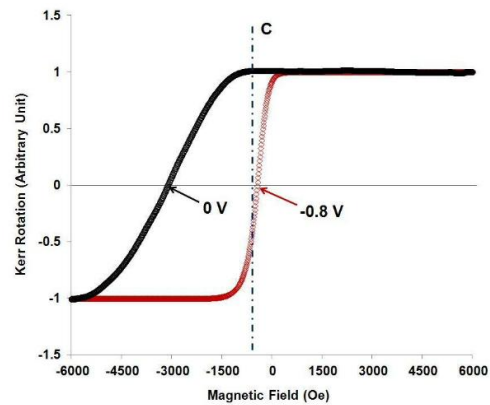

Figure 8: Activation of magnetization switching to a stable magnetization state via gate voltage application

*Conclusion:* The present investigation presents reversible variations in the critical magnetic parameters of a Co/Pt film system by manipulating the induced electrochemical surface charge. Dynamic control of magnetization switching is realized by negatively charging the interface of the ferromagnetic thin film and the electrolyte at various stages of the magnetization reversal process. The possibility of reversibly and dynamically controlling the magnetic properties of ferromagnetic metallic nanostructure, under the assistance of a small bias voltage, at room temperature offers new functionalities to spintronic and magneto- electric devices. This could enable the development of nonvolatile magnetic storage devices with ultra-low power consumption.

*Acknowledgements:* The research at George Washington University (GWU), USA was supported in part by the National Science Foundation (NSF) under Grant No. 1031619. The work at Karlsruhe Institute of Technology (KIT), Germany was supported in part by the Deutsche Forschungsgemeinschaft (DFG) under grant HA 1344/28-1.

______________________